\documentclass{article}

\newtheorem{Claim}{Claim}
\newtheorem{Lemma}{Lemma}
\newtheorem{Theorem}{Theorem}
\newtheorem{Definition}{Definition}
\newtheorem{Corollary}{Corollary}

\newcommand{\comment}[1]{}

\def\proof{\noindent{\bf Proof:~}}
\newcommand{\qed}{\nobreak \ifvmode \relax \else
      \ifdim\lastskip<1.5em \hskip-\lastskip
      \hskip1.5em plus0em minus0.5em \fi \nobreak
      \vrule height0.75em width0.5em depth0.25em\fi}

\newcommand{\ket}[1]{| #1 \rangle}
\newcommand{\bra}[1]{\langle #1|}
\def\H{{\cal H}}
\def\A{{\cal A}}

\begin{document}

\title{Superlinear Advantage for Exact Quantum Algorithms}
\author{Andris Ambainis\thanks{Supported by ESF project 1DP/1.1.1.2.0/09/APIA/VIAA/044 and the European Commission under the FET-Open project QCS (Grant No.~255961), FET-Proactive project QALGO (Grant No. 600700) and ERC Advanced Grant MQC (Grant No. 320731). The conference version of this paper was published in proceedings of STOC'2013.
Part of this work was done at IAS, Princeton, supported by National Science Foundation under agreement No. DMS-1128155. Any opinions, findings and conclusions or recommendations expressed in this material are those of the author(s) and 
do not necessarily reflect the views of the National Science Foundation.}\\
Faculty of Computing \\ University of Latvia\\ Rai\c na~bulv\=aris~19 \\
R\=\i ga, LV-1586, Latvia\\
{E-mail: \tt ambainis@lu.lv}
\date{}
}

\maketitle

\begin{abstract}
A quantum algorithm is exact if, on any input data, it outputs the correct answer with certainty (probability 1). A key question is: how big is the advantage of exact quantum algorithms over their classical counterparts: deterministic algorithms?

We present the first example of a total Boolean function $f(x_1$, $...$, $x_N)$ 
for which exact quantum 
algorithms have superlinear advantage over deterministic algorithms. Any deterministic 
algorithm that computes our function must use $N$ queries but an exact quantum algorithm can 
compute it with $O(N^{0.8675...})$ queries. 

A modification of our function gives a similar 
result for communication complexity: there is a function $f$ which can be computed by an
exact quantum protocol that communicates $O(N^{0.8675...}\log N)$ quantum bits but requires 
$\Omega(N)$ bits of communication for classical protocols.
\end{abstract}

\section{Introduction}

Quantum algorithms can be either studied in the {\em bounded-error} setting (the algorithm
must output the correct answer with probability at least $2/3$, for every input)
or in the {\em exact} setting (the algorithm must output the correct answer with certainty, for every input). 
For the bounded-error case, there are many quantum algorithms that 
are better than classical algorithms (\cite{Shor,Grover,A04a,FGG} and many others).

It is much more difficult to come up with exact quantum algorithms that outperform 
classical algorithms. The requirement that the algorithm's answer must always be correct 
is very constraining: it means that, in the algorithm's final state, we cannot have even very 
small non-zero amplitudes for the basis states that correspond to an incorrect answer. 
Arranging the algorithm's transformations so that this requirement is satisfied for all 
possible inputs has been a very challenging problem.   

We consider computing Boolean functions in the query model. 
Let $Q_E(f)$ ($Q_2(f)$) be the smallest number of queries 
in an exact (bounded-error) quantum algorithm that computes
$f$ and $D(f)$ be the smallest number of queries in a deterministic algorithm that
computes $f$. 

For total Boolean functions, the biggest gap between $Q_E(f)$ and $D(f)$ has been
achieved for the PARITY of $N$ input bits. A modification of Deutsch's algorithm \cite{Deu85} discovered by Cleve et al. \cite{CE+} can compute 
PARITY of 2 input bits exactly with just 1 quantum query. 
This immediately implies that PARITY of $N$ bits can be computed 
with $\lceil N/2 \rceil$ queries. In contrast, deterministic algorithms need $N$ queries to compute PARITY.   

Bigger speedups are known for partial functions.
For example, Brassard and H\o yer \cite{BH} show that 
Simon's algorithm \cite{Sim94} can be made
exact. This gives a partial function $f(x_1, \ldots, x_N)$ with 
$Q_E(f)=O(\log N)$ and $D(f)=\Omega(\sqrt{N})$. 
The value of this function $f(x_1, \ldots, x_N)$, however, is only defined for a very 
small fraction of all inputs $(x_1, \ldots, x_N)$.

Many attempts have been made to come up with exact quantum algorithms for total functions
but the best results have been algorithms that achieve the same separation 
as for the PARITY function: $Q_E(f)=N/2$ vs. $D(f)=N$
(either by using the parity algorithm as a subroutine (e.g. Vasilieva \cite{Vas}) or by different methods (Montanaro et al. \cite{M+})).
\comment{Several authors \cite{Melk,Vas} have given exact quantum algorithms 
that use the exact quantum algorithm for PARITY as a subroutine in an otherwise classical
algorithm. This approach can provide a quantum speedup that is at most a factor of 2. 
More recently, Montanaro et al. \cite{M+} constructed a function $f$ for which $D(f)=4$, $Q_E(f)=2$ 
and there is no 2-query exact quantum algorithm for $f$ that consists of just using the PARITY in an otherwise classical algorithm. This is an interesting result because it provides another approach to constructing exact quantum algorithms. The quantum advantage that is achieved is, however, the same as for PARITY.}

In this paper, we give the first separation between $Q_E(f)$ and $D(f)$ that is more than
a factor of 2. Namely, we obtain 
\[ Q_E(f)=O(D(f)^{0.8675...}) \] for a sequence of functions $f$
(with $D(f)\rightarrow\infty$). 

The sequence of functions is as follows. We start with the function
$NE(x_1, x_2, x_3)$ defined by
\begin{itemize}
\item
$NE(x_1, x_2, x_3)=1$ if $x_i\neq x_j$ for some $i, j\in\{1, 2, 3\}$;
\item
$NE(x_1, x_2, x_3)=0$ if $x_1=x_2=x_3$.
\end{itemize}
We define $NE^1=NE$ and 
\[ NE^d(x_1, \ldots, x_{3^d}) = NE(NE^{d-1}(x_1, \ldots, x_{3^{d-1}}), \]
\[ NE^{d-1}(x_{3^{d-1}+1}, \ldots, x_{2\cdot 3^{d-1}}), 
NE^{d-1}(x_{2\cdot 3^{d-1}+1}, \ldots, x_{3^{d}})) \]
for $d>1$.
This sequence of functions has been known as a candidate for a superlinear separation
between $D(f)$ and $Q_E(f)$ for a long time (it appears that this idea was first mentioned in a 2002 survey by Buhrman and de Wolf \cite{BW02}). 

The reason for that is the relationship between $Q_E(f)$ and the polynomial degree of $f$
by Beals at al. \cite{Beals}.
Let $deg(f)$ be the degree of the unique multilinear polynomial that is equal to $f(x_1,
\ldots, x_N)$. As shown in \cite{Beals}, 
\[ D(f) \geq Q_E(f)\geq deg(f)/2.\]
If we also have $D(f)=deg(f)$ (which is true for many functions $f$), this implies that $Q_E(f)\geq D(f)/2$.

To obtain a bigger gap between $D(f)$ and $Q_E(f)$, we should start with $f$ which has $D(f)>deg(f)$. $NE^d$ has this property.
We have $D(NE)=3$ and $deg(NE)=2$ and one can deduce $D(NE^d)=3^d$ and $deg(NE^d)=2^d$ from that\cite{NS}.
(Kushilevitz \cite{K} and Ambainis \cite{A04} have given 
other constructions of functions with $D(f)>deg(f)$ by taking
a different basis function $f$ instead of $NE$ and iterating it in
the same way.)

Buhrman and de Wolf \cite{BW02} observed that this means the following. If we determine $Q_E(f)$, we will either
get $Q_E(f)=o(D(f))$ or $deg(f)=o(Q_E(f))$ (showing that the degree lower bound on $Q_E(f)$ is not tight). Ambainis \cite{A04} showed that
\[ Q_E(NE^d)\geq Q_2(NE^d) = \Omega(2.121...^d), \]
proving the second of these results. 
For bounded error algorithms, the work on negative adversary bound by H\o yer et al. \cite{HLS} and on span programs by Reichardt and \v Spalek \cite{RS,R09,R11} 
resulted in the conclusion that this bound is optimal: 
$Q_2(NE^d)=\Theta(2.121...^d)$. 

For exact algorithms, there has been no progress,
even though the function $NE^d$ is quite well known. 
In this paper, we provide the first nontrivial exact quantum algorithm for $NE^d$ showing
that 
\[ Q_E(NE^d)=O(2.593...^d)=O(D(NE^d)^{0.8675...}). \]

{\bf Main ideas.}
The main ideas behind our algorithm are as follows. We create an algorithm in which one basis
state has amplitude 1 if $NE^d(x_1, \ldots, x_{3^d})=0$ and an amplitude $\alpha<1$ if 
$NE^d(x_1, \ldots, x_{3^d})=1$. The algorithm is constructed so that $\alpha$ is the same
for all $(x_1, \ldots, x_{3^d})$ with $NE^d(x_1, \ldots, x_{3^d})=1$. The construction is by
induction: we use the algorithm of this type for $NE^{d-1}$ as a subroutine to construct 
the algorithm for $NE^d$. 

Each such induction step decreases the difference between the $NE^d=0$ 
and $NE^d=1$ cases, bringing $\alpha$ closer to 1. To compensate for that, we interleave
the induction steps with a form of quantum amplitude amplification \cite{BHMT} which increases
the difference between $\alpha$ and 1. At the end, we perform the amplitude amplification again, to 
construct an algorithm that perfectly distinguishes between the two cases. 

{\bf Communication complexity.}
As observed by Ronald de Wolf \cite{rdw}, our result also applies to the setting of communication
complexity (in the standard two-party communication model).
Then, we get 
a total function $f(y_1, \ldots, z_N)$ which can be computed by an exact quantum 
protocol that communicates $O(N^{0.8675...} \log N)$ quantum 
bits but requires $\Omega(N)$ bits for classical protocols
in a variety of models (deterministic, bounded-error 
probabilistic and nondeterministic protocol).

Previously, it was known that a classical protocol that communicates $k$ bits
can be converted into a quantum protocol that uses shared entanglement and
$k/2$ quantum bits of communication (via quantum teleportation \cite{Teleport}). 
However, no provable gap of any size between exact 
quantum and deterministic communication complexity was known
for a total function in the case when the quantum protocol does not have
shared entanglement.
(As shown by Buhrman et al. \cite{BCW}, a communication complexity version of the Deutsch-Jozsa problem \cite{DJ}
gives an exponential gap for a partial function.)

\section{Definitions}

We assume familiarity with the standard notions of quantum states and transformations (as described in Nielsen and Chuang \cite{NC} or other textbooks on quantum information). We now briefly define the quantum query model, to synchronize the 
notation with the reader. (For more information on the query model
we refer the reader to the survey by Buhrman and de Wolf \cite{BW02}.)

We assume that the task is 
to compute a Boolean function $f(x_1, \ldots, x_N)$ where $x_1, \ldots, x_N\in\{0, 1\}$.
We consider a Hilbert space $\H$ with basis states $\ket{i, j}$
for $i\in\{0, 1, \ldots, N\}$ and $j\in\{1, \ldots, M\}$ (where $M$ can be 
chosen arbitrarily). We define that a query $Q$ is the following transformation:
\begin{itemize}
\item
$Q\ket{0, j}=\ket{0, j}$;
\item
$Q\ket{i, j}=(-1)^{x_i} \ket{i, j}$ for $i\in\{1, 2, \ldots, N\}$.
\end{itemize}

A quantum query algorithm $\A$ 
consists of a sequence of transformations $U_0$, $Q$, $U_1$,
$\ldots$, $U_{k-1}$, $Q$, $U_k$ where $Q$'s are {\em queries} 
and $U_i$'s (for $i\in\{1, 2, \ldots, k\}$) are arbitrary unitary transformations that do not depend on
$x_1, \ldots, x_N$. The algorithm starts with a fixed starting state
$\ket{\psi_{start}}$ and performs the transformations. This leads to the final
state
\[ \ket{\psi_{final}}= U_k Q U_{k-1} \ldots U_1 Q U_0 \ket{\psi_{start}} .\]
We then measure this state and interpret the result as a binary value $y\in\{0, 1\}$.
(That is, we define some of possible measurement results as corresponding to $y=0$ and
others as corresponding to $y=1$.)

An algorithm $\A$ computes $f(x_1, \ldots, x_N)$ {\em exactly} if, for every 
$x_1, \ldots, x_N\in\{0, 1\}$, the obtained value $y$ is always equal to
$f(x_1, \ldots, x_N)$.

\section{Results and proofs}

\subsection{Results}

Our main result is

\begin{Theorem}
\label{thm:main}
\[ Q_E(NE^d)=O(2.593...^d). \]
\end{Theorem}

Nisan and Szegedy \cite{NS} have shown that $D(NE^d)=3^d$. 
(This follows from the fact that the {\em sensitivity} of $NE^d$ is equal to $3^d$:
$NE^d(0, 0, \ldots, 0)=0$ but, for any input $(x_1$, $\ldots$, $x_N)$ containing exactly one 1,
$NE(x_1, \ldots, x_N)=1$. Hence, any deterministic algorithm must query all $3^d$ variables.)
This means that 
\[ Q_E(NE^d)=O(D(NE^d)^{0.8675...}), \] 
giving a polynomial gap between $D(f)$ and $Q_E(f)$.
One can also show that $R_2(NE^d)$, 
the (bounded error) probabilistic query complexity of $NE^d$, is of the order $\Omega(3^d)$
(since the sensitivity of $NE^d$ on the all-zero input is $3^d$). 
Therefore, we also get the same separation between $Q_E(NE^d)$ and $R_2(NE^d)$.

\subsection{Examples}

We start by giving a few examples which led us to discovering our algorithm. The purpose of this section
is to explain the intuition behind our algorithm. Because of that, we omit the proofs of some statements
(which will be later proven in a more general form in sections \ref{sec:fra}-\ref{sec:proofs}).

{\bf Algorithm 1.}
We consider the following simple algorithm for $NE(x_1, x_2, x_3)$ with
1 query:
\begin{enumerate}
\item
The state space of the algorithm is spanned by basis states $\ket{0}, \ket{1}, \ket{2}, \ket{3}$.
The starting state is 
\[ \ket{\psi_{start}} =
\frac{1}{\sqrt{3}} \ket{1} + \frac{1}{\sqrt{3}} \ket{2} + \frac{1}{\sqrt{3}} \ket{3} .\]
\item
The algorithm performs $U_0=I$, $Q$ and a transformation $U_1$ such that 
$U_1\ket{\psi_{start}}=\ket{0}$, for example,
\[ U_1 = \left( \begin{array}{cccc}
0 & \frac{1}{\sqrt{3}} & \frac{1}{\sqrt{3}} & \frac{1}{\sqrt{3}} \\
\frac{1}{\sqrt{3}} & \frac{1}{\sqrt{3}} & -\frac{1}{\sqrt{3}} & 0 \\
\frac{1}{\sqrt{3}} & 0 & \frac{1}{\sqrt{3}} & -\frac{1}{\sqrt{3}} \\
\frac{1}{\sqrt{3}} & -\frac{1}{\sqrt{3}} & 0 & \frac{1}{\sqrt{3}} 
\end{array} \right) ,\]
with rows and columns numbered in the natural order: $\ket{0}$, $\ket{1}$, $\ket{2}$, $\ket{3}$.
\item
Then, the algorithm measures the state. If the measurement result is 0, 
the algorithm outputs 0. Otherwise, it outputs 1.
\end{enumerate}
We claim that, if $NE(x_1, x_2, x_3)=0$, 
the algorithm always outputs the correct answer 0.
(If $NE(x_1, x_2, x_3)=1$, the algorithm outputs 0 with some probability
and 1 with some probability.)

To see that, we first observe that the final state of this algorithm is 
\[ U_1 Q \ket{\psi_{start}} = \frac{(-1)^{x_1}+(-1)^{x_2}+(-1)^{x_3}}{3} \ket{0}  \]
\[ +\frac{(-1)^{x_1}-(-1)^{x_3}}{3} \ket{1} +
\frac{(-1)^{x_2}-(-1)^{x_1}}{3} \ket{2} +
\frac{(-1)^{x_3}-(-1)^{x_2}}{3} \ket{3}  .\]
If the algorithm outputs 1, 
the amplitude of $\ket{1}, \ket{2}$ or $\ket{3}$ is non-zero and this means
that two of the variables $x_i$ are different.
If $NE(x_1, x_2, x_3)=0$, this never happens.

Moreover, if $NE(x_1, x_2, x_3)=1$, we always have $x_i=x_j$ for exactly one
of pairs $(i, j)$ (where $i, j\in\{1, 2, 3\}, i\neq j$) and $x_i\neq x_j$ for the other two pairs.
Therefore, exactly one of amplitudes of 
$\ket{1}, \ket{2}$ or $\ket{3}$ is zero and the other two amplitudes
are equal to $\pm \frac{2}{3}$. This means that, for any $x_1, x_2, x_3:NE(x_1, x_2, x_3)=1$,
the algorithm outputs 1 with the same probability $2(\frac{2}{3})^2=\frac{8}{9}$. 

{\bf Algorithm 2.}
We would like to transform the algorithm 1 into a form in which it can be used as a
subroutine in a bigger algorithm. To do that, we run algorithm 1 and then perform a sign flip $T$ conditional on the state
being one of the states in which we know that $NE(x_1, x_2, x_3)=1$:
\[ T\ket{0}=\ket{0}, T\ket{1}=-\ket{1}, T\ket{2}=-\ket{2}, T\ket{3}=-\ket{3} .\]
We then perform algorithm 1 in reverse. This results in a 2-query algorithm consisting
of transformations $Q$, $U_1^{-1} T U_1$, $Q$. This algorithm produces the following final state:
\begin{itemize}
\item
If $NE(x_1, x_2, x_3)=0$, $U_1Q\ket{\psi_{start}}$ is $\ket{0}$ or $-\ket{0}$. Hence, 
$T$ has no effect and performing $U^{-1}$ and $Q$ returns the state to $\ket{\psi_{start}}$.
\item
If $NE(x_1, x_2, x_3)=1$, the projection of $U_1 Q \ket{\psi_{start}}$ to the subspace
spanned by $\ket{1}$, $\ket{2}$, $\ket{3}$ is always of the same length $\sqrt{8/9}$ 
(square root of the probability that Algorithm 1 outputs 1). In other words, $T$ always 
flips the sign on a part of the state with the same length $\sqrt{8/9}$. 
It can be shown\footnote{The proof is a particular case of Case 2 of Lemma \ref{lem:iterate}.} 
that this implies 
\[ Q U_1^{-1} T U_1 Q \ket{\psi_{start}} = - \frac{7}{9} 
\ket{\psi_{start}} + \ket{\psi^{\perp}} \]
for some $\ket{\psi^{\perp}} \perp \ket{\psi_{start}}$ that 
depends on the values of $x_1, x_2, x_3$.
\end{itemize}
We now see that the amplitude of $\ket{\psi_{start}}$
in the final state depends only on the value of $NE(x_1, x_2, x_3)$ (it is 
$-\frac{7}{9}$ if $NE=1$ and 1 if $NE=0$).
This is important because it allows to use the algorithm as a subroutine in the next step.

{\bf Algorithm 3.} Next, we construct an algorithm for $NE^2(y_1$, $\ldots$, $y_9)$
by taking Algorithm 1 and 
substituting copies of Algorithm 2 computing
\[ NE(y_1, y_2, y_3), NE(y_4, y_5, y_6), NE(y_7, y_8, y_9) \] 
instead of queries to $x_1, x_2, x_3$.

This substitution is carried out as follows:
We create 3 copies of the working space of Algorithm 2, 
with $\ket{\psi_{start, i}}$, $i\in\{1, 2, 3\}$ as the starting states.
We also add an extra basis state $\ket{0}$ which is orthogonal to all 3 
copies of the workspace. Algorithm 3 works as follows:
\begin{enumerate}
\item
Algorithm's starting state is 
\[ \ket{\psi_{start}} =
\frac{1}{\sqrt{3}} \ket{\psi_{start, 1}} + 
\frac{1}{\sqrt{3}} \ket{\psi_{start, 2}} + 
\frac{1}{\sqrt{3}} \ket{\psi_{start, 3}} .\]
\item
The algorithm runs 3 copies of Algorithm 2 in parallel on the 3 copies of its 
workspace, for $NE(y_1, y_2, y_3)$, $NE(y_4, y_5, y_6)$, $NE(y_7, y_8, y_9)$. 
\item
Then, it performs the transformation $U_2$ 
on the subspace spanned by $\ket{0}$ and $\ket{\psi_{start, i}}$:
\[ U_2\ket{\psi_{start, 1}} = \frac{1}{\sqrt{3}} \ket{0} + 
\frac{1}{\sqrt{3}} \ket{\psi_{start, 1}} - \frac{1}{\sqrt{3}} \ket{\psi_{start, 2}} ,\]
\[  U_2\ket{\psi_{start, 2}} = \frac{1}{\sqrt{3}} \ket{0} + 
\frac{1}{\sqrt{3}} \ket{\psi_{start, 2}} - \frac{1}{\sqrt{3}} \ket{\psi_{start, 3}} ,\]
\[ U_2\ket{\psi_{start, 3}} = \frac{1}{\sqrt{3}} \ket{0} + \frac{1}{\sqrt{3}} 
\ket{\psi_{start, 3}} - \frac{1}{\sqrt{3}} \ket{\psi_{start, 1}} \]
which maps $\ket{\psi_{start}}$ to $\ket{0}$.
\end{enumerate}
This algorithm has the following property:

\begin{Claim}
\label{cl:ort}
If $NE^2(x_1, \ldots, x_9)=0$,
the final state of Algorithm 3 is orthogonal to all $\ket{\psi_{start, i}}$.
\end{Claim}

\proof
$NE^2(x_1, \ldots, x_9)=0$ means that
$NE(x_1, x_2, x_3)$, $NE(x_4, x_5, x_6)$, $NE(x_7, x_8, x_9)$
are either all equal to 0 or all equal to 1.
If they are all 0, Algorithm 2 maps each of states $\ket{\psi_{start, i}}$ 
to itself. Hence, the state $\ket{\psi_{start}}$ is mapped to itself 
by Algorithm 2 and to $\ket{0}$ by $U_2$.

If they are all 1, Algorithm 2 maps $\ket{\psi_{start}}$ to 
\[ -\frac{7}{9\sqrt{3}} \ket{\psi_{start, 1}}  
-\frac{7}{9\sqrt{3}} \ket{\psi_{start, 2}}  
-\frac{7}{9\sqrt{3}} \ket{\psi_{start, 3}} + \ket{\psi^{\perp}} \]
\[ = -\frac{7}{9} \ket{\psi_{start}} + \ket{\psi^{\perp}} .\]
where $\ket{\psi^{\perp}}$ is orthogonal to all of $\ket{\psi_{start, i}}$.
$U_2$ then maps this state to $-\frac{7}{9} \ket{0} +\ket{\psi^{\perp}}$
(because $U_2\ket{\psi_{start}}=\ket{0}$ and $U_2\ket{\psi^{\perp}}=\ket{\psi^{\perp}}$).
The resulting state $-\frac{7}{9} \ket{0} +\ket{\psi^{\perp}}$ 
is orthogonal to all $\ket{\psi_{start, i}}$.
\qed

In the case if $NE^2(x_1, \ldots, x_9)=1$, we again have the property that
the projection of the final state of Algorithm 3 
to the subspace spanned by all $\ket{\psi_{start, i}}$ is always of the same length -
independent of $x_1, \ldots, x_9$. (Again, this is a particular case of Case 2 of Lemma \ref{lem:iterate} and it
follows from Algorithm 2 having similar property. And, again, we omit the proof.)

{\bf Algorithms 4, 5, etc.}
We can repeat this construction, by taking Algorithm 3 and transforming it into a form in which it can be used as a subroutine 
(similarly how we obtained Algorithm 2 from Algorithm 1). We can then take the resulting Algorithm 4 and use it as
a subroutine in an algorithm computing $NE^3(x_1, \ldots, x_{27})$ and so on.

\comment{
We can iterate the algorithm further, in a similar way. Each next iteration
degrades the performance and makes the difference between the $NE^d=0$ and the $NE^d=1$
cases smaller. 

To compensate for that, we combine the iteration with a form of
quantum amplitude amplification \cite{BHMT} which increases the number of queries
but also improves the difference between the $NE^d=0$ and the $NE^d=1$
cases.}

\subsection{Framework}
\label{sec:fra}

Next, we develop these ideas in a more general form.
Similarly to Algorithm 2, we construct algorithms which produce a final state
in which one amplitude takes one of two values - depending 
on $f(x_1, \ldots, x_N)$:

\begin{Definition}
\label{def:p}
Let $p\in[-1, 1]$. A quantum query algorithm $\A$ $p$-computes 
a function $f(x_1, \ldots, x_N)$ if, for some $\ket{\psi_{start}}$:
\begin{enumerate}
\item[(a)]
$\A\ket{\psi_{start}} = \ket{\psi_{start}}$ whenever $f(x_1, \ldots, x_N)=0$;
\item[(b)]
if $f(x_1, \ldots, x_N)=1$, then
\[ \A\ket{\psi_{start}} = p\ket{\psi_{start}}+\sqrt{1-p^2}\ket{\psi} \]
for some $\ket{\psi}:\ket{\psi}\perp\ket{\psi_{start}}$
(where $\ket{\psi}$ may depend on $x_1, \ldots, x_N$).
\end{enumerate}
\end{Definition}

We note that $p$-computing a function becomes easier when $p$ increases. The easiest 
case is $p=1$ when any function can be 1-computed in a trivial way by performing 
the identity transformation $I$ on $\ket{\psi_{start}}$. For $p=0$, an algorithm
$\A$ that 0-computes $f$ is also an exact algorithm for $f$ in the usual sense
because we can measure whether the final state of $\A$  is
$\ket{\psi_{start}}$ or orthogonal to $\ket{\psi_{start}}$ and output 0 in the first case
and 1 in the second case. 

For $p=-1$, $p$-computing $f$ means that $\A\ket{\psi_{start}} = \ket{\psi_{start}}$
whenever $f=0$ and $\A\ket{\psi_{start}} = -\ket{\psi_{start}}$. This is the same 
transformation as the query black box. Hence, if we consider the iterated function 
\[ f(f(y_1, \ldots, y_N), f(y_{N+1}, \ldots, y_{2N}), \ldots) \]
we can obtain an algorithm for (-1)-computing it by taking the algorithm $\A$
for (-1)-computing $f(x_1, x_2, \ldots, x_N)$ and, instead of queries to $x_i$, running
$\A$ to (-1)-compute $f(y_{(i-1)N-1}, \ldots, y_{iN})$. Thus, an algorithm for (-1)-computing
$f$ with $k$ queries immediately implies an algorithm for $f^d$ with $k^d$ queries. 
(In contrast, if we had to iterate an exact algorithm for $f$ in the usual sense,
we would have to run it twice: in the forward direction to compute $f$ and 
in reverse to uncompute the unnecessary extra information. As a result, we would obtain 
an exact algorithm for $f^d$ with $(2k)^d$ queries.)

Because of this property, our goal is to obtain algorithms
for (-1)-computing $NE^d$ for some $d$. We will use algorithms for $p$-computing 
$NE^d$ with $p>-1$ as stepping stones in our construction.

We also have

\begin{Lemma}
\label{lem:transform}
If an algorithm $\A$ $p$-computes $f$ with $k$ queries, there is an 
algorithm $\A'$ that $p'$-computes $f$ with $k$ queries, for any $p'>p$.
\end{Lemma}

\proof
We enlarge the state space of the algorithm by adding a new basis state 
$\ket{0, j}$ which is left unchanged by queries $Q$. We define $\A'$ by extending all 
$U_i$ to the enlarged state space by defining $U_i\ket{0, j}=\ket{0, j}$
and change the starting state to 
$\ket{\psi'_{start}} = \cos \alpha \ket{\psi_{start}} 
+ \sin  \alpha \ket{0, j}$.

If $f(x_1, \ldots, x_N)=0$, we have $\A\ket{\psi_{start}}=\ket{\psi_{start}}$ and, hence, 
\[ \A'\ket{\psi'_{start}}=\ket{\psi'_{start}}. \]

If $f(x_1, \ldots, x_N)=1$, then 
\[ \A'\ket{\psi'_{start}}= \cos \alpha \A\ket{\psi_{start}} + \sin \alpha \ket{0, j}  \]
\[ = p \cos \alpha \ket{\psi_{start}} + \sqrt{1-p^2}\cos \alpha \ket{\psi^{\perp}}+ \sin \alpha \ket{0, j}
\]
where $\ket{\psi^{\perp}}$ is perpendicular to both $\ket{\psi_{start}}$ and $\ket{0, j}$.
Hence, 
\[ \bra{\psi'_{start}} \A'\ket{\psi'_{start}} = p \cos^2 \alpha + \sin^2 \alpha \]
and we have
$\A'\ket{\psi'_{start}}=p'\ket{\psi'_{start}}+\sqrt{1-(p')^2}\ket{\psi'}$
($\ket{\psi'} \perp \ket{\psi'_{start}}$) for
$p'= p \cos^2 \alpha + \sin^2 \alpha$.
By varying $\alpha$ over the interval $[0, \frac{\pi}{2}]$, we can achieve any value
of $p'$ between $p'=p$ (for $\alpha=0$) and $p'=1$ (for $\alpha=\frac{\pi}{2}$).
\qed

If we have an algorithm $\A$ that $p$-computes $NE^{i-1}$, we can use it to build an
algorithm $\A'$ that $p'$-computes $NE^i$, through the following lemma.

\begin{Lemma}
\label{lem:iterate}
If an algorithm $\A$ $p$-computes $NE^{d-1}$ with $k$ queries, there is an 
algorithm $\A'$ that $p'$-computes $NE^d$ with $2k$ queries, for 
$p'= 1 - \frac{4(1-p)^2}{9}$.
\end{Lemma}

\proof
In section \ref{sec:proofs}.
\qed

Applying Lemma \ref{lem:iterate} results in an algorithm with $p'>p$. 
Applying it several times degrades $p$ even further (that is, $p$ becomes closer and
closer to 1). To compensate for that, we have the following lemma for improving $p$ (which follows by adapting quantum amplitude amplification \cite{BHMT} to our setting).

\begin{Lemma}
\label{lem:amplify1}
If an algorithm $\A$ $p$-computes a function $f$ with $k$ queries, for $p=\cos \alpha$, 
there is an algorithm $\A'$ that $p'$-computes $f$ with $ck$ queries, for 
$p'= \cos c \alpha$.
\end{Lemma}

\proof
In section \ref{sec:proofs}.
\qed

As a special case of Lemma \ref{lem:amplify1}, we obtain 

\begin{Corollary}
\label{lem:amplify}
If an algorithm $\A$ $p$-computes $NE^d$ with $k$ queries, there is an 
algorithm $\A'$ that $p'$-computes $NE^d$ with $2k$ queries, for 
$p'= 2p^2 -1$.
\end{Corollary}

\proof
We set $c=2$ in Lemma \ref{lem:amplify1}. Then,
\[ p'=\cos 2 (\arccos p) = 2p^2 -1. \]
\qed

\subsection{Algorithm for $NE^d$}

It is easy to see that

\begin{Lemma}
If there is an algorithm $\A$ that $(-1)$-computes $NE^t$ with $k$ queries,
there is an algorithm $\A_l$ that $(-1)$-computes $NE^{tl}$ with $k^l$ queries 
for all $l\geq 1$.
\end{Lemma}

\proof
By induction. The algorithm $\A$ itself forms the base case, for $l=1$.

For the inductive case, we can obtain the function $NE^{tl}$ by taking
the function $NE^{t(l-1)}$ and, instead of each variable, 
substituting a function $NE^t$ on a block of $3^t$ new variables.
Therefore, we can take the algorithm $\A_{l-1}$ that computes
$NE^{t(l-1)}$ with $k^{l-1}$ queries and,
instead of each query, substitute the algorithm $\A$ that performs 
$\A\ket{\psi}=\ket{\psi}$ if $NE^t=0$ for the corresponding 
group of $3^t$ variables and $\A\ket{\psi}=-\ket{\psi}$ if $NE^t=1$. 

In the resulting algorithm $\A_l$, each of $k^{l-1}$ queries of $\A_l$
is replaced by a sequence of transformations that involves $k$ queries.
Therefore, the total number of queries for $\A_l$ is $k^{l-1} k = k^l$.
\qed

Moreover, we get that $Q_E(NE^{tl})\leq k^l$, since an algorithm that $(-1)$-computes
$NE^{tl}$ can be transformed into an algorithm that 0-computes $NE^{tl}$, with no
increase in the number of queries (Lemma \ref{lem:transform}) and an algorithm for 0-computing $f$ can be used to compute $f$ exactly. Therefore,
we have

\begin{Corollary}
\label{cor:main}
If there exists an algorithm $\A$ that $(-1)$-computes $NE^t$ with $k$ queries,
then $Q_E(NE^d)=O(k^{d/t})$.
\end{Corollary}

It remains to find a base algorithm that $(-1)$-computes $NE^d$ with less than $3^d$ queries. We give two such constructions.

{\bf Construction 1:}
\begin{enumerate}
\item
$NE^0(x_1)=x_1$ can be $(-1)$-computed with 1 query
by just performing a query $\ket{1}\rightarrow (-1)^{x_1} \ket{1}$.
\item
Applying Lemma \ref{lem:iterate} with $p=-1$ gives that $NE^1$ can be $(-\frac{7}{9})$-computed with 2 queries.
\item
Applying Lemma \ref{lem:iterate} with $p=-\frac{7}{9}$ gives that $NE^2$ can be
$p'$-computed with 4 queries for $p'=-\frac{295}{729}$.
\item
Applying Lemma \ref{lem:transform} gives that $NE^2$ can be 0-computed with 4 queries.
\item
Applying Corollary \ref{lem:amplify} with $p=0$ gives that $NE^2$ can be $(-1)$-computed with 8 queries. 
\end{enumerate}

By Corollary \ref{cor:main}, this means that 
$Q_E(NE^d)=O(8^{d/2})=O(2.828...^d)$.
This result can be improved by using a more complicated base construction with a larger number of steps. 

{\bf Construction 2:}
\begin{enumerate}
\item
Start with an algorithm that 
$(-\frac{295}{729})$-computes $NE^2$ with 4 queries (from Construction 1).
\item
Applying Lemma \ref{lem:iterate} with $p=-\frac{295}{729}$ gives that $NE^3$ can be
$p'$-computed with 8 queries for $p'=\frac{588665}{4782969}=0.123075...$.
\item
Applying Corollary \ref{lem:amplify} gives that $NE^3$ can be $p'$-computed with 
16 queries for $p'=-0.969704...$.
\item
Applying Lemma \ref{lem:iterate} three times gives that $NE^6$ can be $p'$-computed with 
128 queries for $p'=0.223874...$.
\item
Applying Corollary \ref{lem:amplify} gives that $NE^6$ can be $p'$-computed with 
256 queries for $p'=-0.8997602...$.
\item
Applying Lemma \ref{lem:iterate} two times gives that $NE^8$ can be $p'$-computed with 
1024 queries for $p'=-0.14353...$.
\item
Applying Lemma \ref{lem:transform} gives that $NE^8$ can be also 0-computed with 1024 queries and applying 
Corollary \ref{lem:amplify} gives that $NE^8$ can be $(-1)$-computed with
2048 queries,
\end{enumerate}
This means that $Q_E(NE^d)=O(2048^{d/8})= O(2.593...^d)$. 
Computer experiments \cite{Balodis} show that this is the best result
that can be achieved by combining Lemmas \ref{lem:transform} and \ref{lem:amplify1}
and Corollary \ref{lem:amplify}, as long as $d\leq 1000$
and all intermediate algorithms have $p<0.99999$.
We suspect that nothing can be gained by using larger $d$ or $p$.

Another possibility would be to construct an algorithm for $p$-computing $NE^d$ 
by other means. In this direction, 
semidefinite optimization \cite{Iraids} shows that our algorithms for 
$NE^1$ and $NE^2$ are optimal: $NE^1$ cannot be $p$-computed with 2 queries
for $p<-\frac{7}{9}$ and $NE^2$ cannot be $p$-computed with 4 queries
for $p<-\frac{295}{729}$. We do not know whether the algorithms for $NE^d$, $d>2$ are
optimal.

\subsection{Proofs of Lemmas}
\label{sec:proofs}

\noindent
{\bf Lemma 2.} {\em If an algorithm $\A$ $p$-computes $NE^{d-1}$ with $k$ queries, there is an 
algorithm $\A'$ that $p'$-computes $NE^d$ with $2k$ queries, for 
$p'= 1 - \frac{4(1-p)^2}{9}$.}

\medskip

\proof
We assume that the algorithm $\A$ consists of transformations $U_0$, $Q$, $U_1$,
$\ldots$, $U_{k-1}$, $Q$, $U_k$, in a Hilbert space 
$\H$ with basis states $\ket{i, j}$, $i\in\{0, 1, \ldots, 3^{d-1}\}$, $j\in\{1, \ldots, M\}$.
Let $\ket{\psi_{start}}$ be the starting state of $\A$.

We consider a Hilbert space $\H'$ with basis states $\ket{i, j}$, 
$i\in\{0, 1, \ldots, 3^{d}\}$, $j\in\{1, \ldots, 3M\}$.
In this Hilbert space, we can compute 
$NE^{d-1}(x_1, \ldots, x_{3^{d-1}})$, 
$NE^{d-1}(x_{3^{d-1}+1}, \ldots, x_{2\cdot 3^{d-1}})$, 
$NE^{d-1}(x_{2\cdot 3^{d-1}+1}, \ldots, x_{3^{d}})$ in parallel, in the following
way. 

Let $\H_l$ ($l\in\{1, 2, 3\}$) be the subspace spanned by the 
basis states 
\[ \ket{0, j},
j\in\{(l-1)M+1, \ldots, l M\} \] 
and 
\[ \ket{i, j}, 
i\in\{(l-1)3^{d-1}+1, \ldots, l 3^{d-1}\}, j\in\{1, \ldots, M\}. \]  
We have an isomorphism $V_l:\H_l\rightarrow \H$ defined by 
\[ V_l\ket{0, (l-1)M+j}= \ket{0, j}, \]
\[ V_l \ket{(l-1)3^{d-1}+i, j}=\ket{i, j} .\]
We define $U^{(l)}_i=(V_l)^{\dagger} U_i V_l$ and 
$\ket{\psi^{(l)}_{start}}= (V_l)^{\dagger} \ket{\psi_{start}}$.
Then, the sequence of transformations 
\[ U^{(l)}_0, Q, U^{(l)}_1,
\ldots, U^{(l)}_{k-1}, Q, U^{(l)}_k \]
with the starting state 
$\ket{\psi^{(l)}_{start}}$ 
is a quantum algorithm $p$-computing the function $NE^{d-1}(x_{(l-1) 3^{d-1}+1}, 
\ldots, x_{l\cdot 3^{d-1}})$. 

Let $U'_i$ be a transformation which is equal to $U^{(l)}_i$ on $\H_l$, for
each $l\in\{1, 2, 3\}$. Then, the sequence of transformations
$U'_0$, $Q$, $U'_1$, $\ldots$, $U'_{k-1}$, $Q$, $U'_k$ can be used to $p$-compute
any of $NE^{d-1}((l-1) x_{3^{d-1}+1}, \ldots, x_{l\cdot 3^{d-1}})$, 
for $l\in\{1, 2, 3\}$, depending
on the starting state. Let
\[ V = U'_k Q U'_{k-1} \cdots U'_1 Q U'_0 \]
be the product of these transformations.

Let $T$ be the unitary transformation defined by:
\begin{itemize}
\item
$T\ket{\psi'_{start}}=\ket{\psi'_{start}}$ for $\ket{\psi'_{start}}=\sum_{l=1}^3 \frac{1}{\sqrt{3}} 
\ket{\psi^{(l)}_{start}}$;
\item
$T\ket{\psi}=-\ket{\psi}$ for any $\ket{\psi}$ 
that is a linear combination of $\ket{\psi^{(1)}_{start}}$, 
$\ket{\psi^{(2)}_{start}}$, $\ket{\psi^{(3)}_{start}}$ and 
satisfies $\ket{\psi}\perp \ket{\psi'_{start}}$;
\item
$T\ket{\psi}=\ket{\psi}$ for any $\ket{\psi}$ that is perpendicular to all of 
$\ket{\psi^{(1)}_{start}}$, $\ket{\psi^{(2)}_{start}}$, $\ket{\psi^{(3)}_{start}}$.
\end{itemize}

We take the algorithm $\A'$ which performs the transformation
$\A'=V^{-1} T V$, on the starting state $\ket{\psi'_{start}}$. 
We claim that this algorithm 
$p'$-computes the function $NE^d$.

To prove it, we consider two cases.

{\bf Case 1:}
$NE^d(x_1, \ldots, x_{3^d})=0$.

This means that we have one of two subcases.

{\bf Case 1a:}
$NE^{d-1}(x_{(l-1) 3^{d-1}+1}, \ldots, x_{l\cdot 3^{d-1}}) = 
0$ for all $l\in\{1, 2, 3\}$.

Then, 
given a starting state
$\ket{\psi^{(l)}_{start}}$, $V$ performs the algorithm for $p$-computing 
\[ NE^{d-1}(x_{(l-1) 3^{d-1}+1}, \ldots, x_{l\cdot 3^{d-1}}) = 0. \]
This means that $V\ket{\psi^{(l)}_{start}}=\ket{\psi^{(l)}_{start}}$
for all $l\in\{1, 2, 3\}$ and, hence, $V\ket{\psi'_{start}}=\ket{\psi'_{start}}$.
Since $T\ket{\psi'_{start}}=\ket{\psi'_{start}}$ (by the definition of $T$),
we get that $\A'\ket{\psi'_{start}}=\ket{\psi'_{start}}$.

{\bf Case 1b:}
$NE^{d-1}((l-1) x_{3^{d-1}+1}, \ldots, x_{l\cdot 3^{d-1}}) = 
1$ for all $l\in\{1, 2, 3\}$.

Then, 
\[ V\ket{\psi^{(l)}_{start}}=p \ket{\psi^{(l)}_{start}}+ \sqrt{1-p^2}\ket{\psi^{(l)}} \]
where $\ket{\psi^{(l)}}\in \H_l$ and $\ket{\psi^{(l)}}\perp \ket{\psi^{(l)}_{start}}$.
Trivially, we also have $\ket{\psi^{(l)}} \perp \ket{\psi^{(l')}_{start}}$ for $l \neq l'$
(since $\ket{\psi^{(l)}}\in \H_l$, $\ket{\psi^{(l')}_{start}}\in \H_{l'}$ and 
$\H_l \perp \H_{l'}$). This means that applying $T$ to
\[ V \ket{\psi'_{start}} = \frac{1}{\sqrt{3}} \sum_{l=1}^3 p \ket{\psi^{(l)}_{start}} +
\frac{1}{\sqrt{3}} \sqrt{1-p^2} \sum_{l=1}^3 \ket{\psi^{(l)}} .\]
does not change the state because the first component is equal to
$p\ket{\psi'_{start}}$ and the second component is orthogonal to all 
$\ket{\psi^{(l)}_{start}}$. 
Hence, 
\[ V^{-1} T V \ket{\psi'_{start}} = V^{-1} V \ket{\psi'_{start}} =
\ket{\psi'_{start}} .\]

{\bf Case 2:}
$NE^d(x_1, \ldots, x_{3^d})=1$.

In this case, we have to prove 
\[ \bra{\psi'_{start}} V^{-1} T V \ket{\psi'_{start}} = p'.\]
We express 
\[ V\ket{\psi'_{start}}=\alpha \ket{\psi_+}+\beta \ket{\psi_-} \]
where $T\ket{\psi_+}=\ket{\psi_+}$ and $T\ket{\psi_-}=-\ket{\psi_-}$
and $\alpha, \beta$ satisfy $|\alpha|^2+|\beta|^2 =1$.
Then, 
\[ \bra{\psi'_{start}} V^{-1} T V \ket{\psi'_{start}} = |\alpha|^2 -|\beta|^2 =
1- 2 |\beta|^2 .\]
To calculate $\beta$, we consider two subcases:

{\bf Case 2a:}
$NE^{d-1}(x_{(l-1) 3^{d-1}+1}, \ldots, x_{l\cdot 3^{d-1}}) = 1$ for exactly
one of $l\in\{1, 2, 3\}$.

Without loss of generality, we assume that this is $l=1$. Then, 
\[ V \ket{\psi'_{start}} =
  \frac{1}{\sqrt{3}} \left( p \ket{\psi^{(1)}_{start}} +\sqrt{1-p^2} \ket{\psi^{(1)}} + \ket{\psi^{(2)}_{start}} + \ket{\psi^{(3)}_{start}} \right) .\]
We have
\[ \beta \ket{\psi_-} = \frac{2p-2}{3\sqrt{3}} \ket{\psi^{(1)}_{start}} + 
\frac{1-p}{3\sqrt{3}} \ket{\psi^{(2)}_{start}} 
+ \frac{1-p}{3\sqrt{3}} \ket{\psi^{(3)}_{start}} , \]
\[ |\beta|^2 = \frac{(2p-2)^2}{27} + 2 \frac{(1-p)^2}{27} = \frac{2(1-p)^2}{9} .\] 

{\bf Case 2b:}
$NE^{d-1}(x_{(l-1) 3^{d-1}+1}, \ldots, x_{l\cdot 3^{d-1}}) = 1$ for exactly
two of $l\in\{1, 2, 3\}$.

Without loss of generality, we assume that these are $l=1$ and $l=2$. Then,
\[ V \ket{\psi'_{start}}= \frac{1}{\sqrt{3}} \left( p \ket{\psi^{(1)}_{start}} +\sqrt{1-p^2} \ket{\psi^{(1)}} \right. \]
\[ \left. + p \ket{\psi^{(2)}_{start}} +\sqrt{1-p^2} \ket{\psi^{(2)}} + \ket{\psi^{(3)}_{start}} \right) .\]
We have 
\[ \beta \ket{\psi_-} = \frac{p-1}{3\sqrt{3}} \ket{\psi^{(1)}_{start}} + 
\frac{p-1}{3\sqrt{3}} \ket{\psi^{(2)}_{start}} 
+ \frac{2-2p}{3\sqrt{3}} \ket{\psi^{(3)}_{start}}  \]
and, similarly to the previous case, we get $|\beta|^2 = \frac{2(1-p)^2}{9}$.
\qed
\medskip 

\noindent
{\bf Lemma 3.} {\em If an algorithm $\A$ $p$-computes a function $f$ with $k$ queries, for $p=\cos \alpha$, 
there is an algorithm $\A'$ that $p'$-computes $f$ with $ck$ queries, for 
$p'= \cos c \alpha$.}
\medskip 

\proof [of Lemma \ref{lem:amplify1}]
Let $\ket{\psi_{start}}$ be the starting state of $\A$. 
Let $T$ be the transformation defined by $T\ket{\psi_{start}}=\ket{\psi_{start}}$
and $T\ket{\psi}=-\ket{\psi}$ for all $\ket{\psi}:\ket{\psi}\perp\ket{\psi_{start}}$. 

The new algorithm $\A'$ has the same starting state $\ket{\psi_{start}}$ and consists of  transformations $V_1, T, V_2, T, \ldots, T, V_c$
where $V_i=\A$ for odd $i$ and $V_i=\A^{-1}$ for even $i$. Since $\A$ and $\A^{-1}$ both use $k$ queries and $T$ can be performed with
no queries, the new algorithm $\A'$ uses $ck$ queries. 

If $f(x_1, \ldots, x_N)=0$, then (by definition \ref{def:p}), $\A\ket{\psi_{start}}=\ket{\psi_{start}}$. 
Since this also means $\A^{-1}\ket{\psi_{start}}=\ket{\psi_{start}}$, we get that
$\A'\ket{\psi_{start}}=\ket{\psi_{start}}$. 

In the $f(x_1, \ldots, x_N)=1$ case, we have
\begin{equation}
\label{eq:forward} \A\ket{\psi_{start}} = \cos \alpha \ket{\psi_{start}}+ \sin\alpha \ket{\psi_1} 
\end{equation}
for some $\ket{\psi_1}\perp\ket{\psi_{start}}$.
If $c$ is odd, we have $\A' = (\A T \A^{-1} T)^{(c-1)/2} \A$.
This is a particular case of the standard setting of quantum amplitude amplification \cite{BHMT} 
and the analysis of Brassard et al. \cite{BHMT} implies that 
\[ \A\ket{\psi_{start}} = \cos c \alpha \ket{\psi_{start}}+ \sin c\alpha \ket{\psi_1} .\]
The even $c$ case can be analyzed in a similar way. We observe that (\ref{eq:forward}) implies 
\[ \A^{-1}\ket{\psi_{start}} = \cos \alpha \ket{\psi_{start}}+ \sin\alpha \ket{\psi_2} \]
for some $\ket{\psi_2}\perp\ket{\psi_{start}}$ (because $\A$ is unitary and, therefore, 
the inner product between $\ket{\psi_{start}}$ and $\A\ket{\psi_{start}}$ must be equal to
the inner product between $\A^{-1}\ket{\psi_{start}}$ and $\A^{-1}\A\ket{\psi_{start}}=\ket{\psi_{start}}$).

We can express $\A' = T^{-1} (T \A^{-1}T\A)^{c/2}= T (T \A^{-1}T\A)^{c/2}$. The transformation $T \A^{-1}T\A$ can
be viewed as a product of two reflections $T$ and $\A^{-1}T\A$. Each of these reflections leaves
a certain state ($\ket{\psi_{start}}$ for $T$ and $\A^{-1}\ket{\psi_{start}}$ for $\A^{-1}T\A$) unchanged and
maps any $\ket{\psi}$ which is orthogonal to this state to $-\ket{\psi}$. 

We consider the two dimensional subspace $\H_2$ spanned by $\ket{\psi_{start}}$ and $A^{-1}\ket{\psi_{start}}$.
Both $T$ and $\A^{-1}T\A$ map this subspace to itself. (For $T$, the subspace $\H_2$ is spanned by
$\ket{\psi_{start}}$ and a state $\ket{\psi^{\perp}}\perp\ket{\psi_{start}}$. $T$ maps $\ket{\psi_{start}}$
to itself and $\ket{\psi^{\perp}}$ to $-\ket{\psi^{\perp}}$. Therefore, $T(\H_2)=\H_2$.
For $\A^{-1}T\A$, the proof is similar, with $\A^{-1}\ket{\psi_{start}}$ instead of $\ket{\psi_{start}}$.)
Since the starting state of $\A'$ is $\ket{\psi_{start}}\in \H_2$, this means that algorithm's state always 
stays within $\H_2$.

A composition of two reflections in a 2-dimensional subspace with respect to vectors ($\ket{\psi_{start}}$ and $A\ket{\psi_{start}}$) 
that are at an angle $\alpha$ one to another is equal to a rotation by an angle $2\alpha$ (as used in ``two reflections" analysis
of Grover's algorithm \cite{Aharonov} and amplitude amplification \cite{BHMT}) Repeating such a sequence of two reflections $c/2$ times leads
to a state that is at an angle $2(c/2)\alpha=c\alpha$ with a starting state, i.e. to a state
\[ (T \A^{-1}T\A)^{c/2}\ket{\psi_{start}} = \cos c\alpha \ket{\psi_{start}}+ \sin c\alpha \ket{\psi_3} \]
for some $\ket{\psi_3}\perp \ket{\psi_{start}}$. We then have
\[ \A' \ket{\psi_{start}} = T (T \A^{-1}T\A)^{c/2}\ket{\psi_{start}}   =\cos c\alpha \ket{\psi_{start}}- \sin c\alpha \ket{\psi_3} .\]
\qed

\subsection{Implications for communication complexity}

We consider the problem of computing a function 
\[ f(y_1, \ldots, y_N, z_1, \ldots, z_N) \] in the communication complexity setting
in which one party (Alice) holds $y_1, \ldots, y_N$ and another party (Bob)
holds $z_1$, $\ldots$, $z_N$. The task is to compute $f(y_1, \ldots, z_N)$
with the minimum communication. (For a review on quantum communication complexity,
see Buhrman et al. \cite{BC+}.)

A quantum communication protocol is exact if, after communicating $k$ qubits 
(for some fixed $k$), both parties output answers that
are always equal to $f(y_1, \ldots, z_N)$.
The communication complexity counterpart of Theorem \ref{thm:main} is 
the following theorem (due to Ronald de Wolf \cite{rdw}):
\begin{Theorem}
\label{thm:rdw}
\cite{rdw}
There exists $f(y_1, \ldots, y_N, z_1, \ldots, z_N)$ such that
\begin{enumerate}
\item
$f$ can be computed exactly by a quantum protocol with communication of $O(N^{0.8675...}
\log N)$
quantum bits;
\item
Any deterministic protocol (or bounded-error probabilistic protocol
or nondeterministic protocol) computing $f$ communicates at least $\Omega(N)$ bits.
\end{enumerate}
\end{Theorem}

\proof
Let $N=3^d$. We define 
\[ f(y_1, \ldots, z_N) = NE^d(y_1 \wedge z_1, \ldots, y_N \wedge z_N) .\]
As shown by Buhrman et al. \cite{BCW}, existence of a $T$ query quantum algorithm for 
$g(x_1, \ldots, x_N)$ implies the existence of a quantum protocol 
for $g(y_1\wedge z_1, \ldots, y_N \wedge z_N)$ that 
communicates $O(T \log N)$ quantum bits. (Alice runs the query algorithm
and implements each query through $O(\log N)$ quantum bits of communication
with Bob.) This implies the first part of the theorem.

The second part follows by a reduction from the set disjointness problem \cite{ks,razborov}.
We define $DISJ(y_1, \ldots, z_N)= 1$ if the sets $\{i:y_i=1\}$
and $\{i:z_i=1\}$ are disjoint and $DISJ(y_1, \ldots, z_N)= 0$ if these two sets are not
disjoint. This is equivalent to
\[ DISJ(y_1, \ldots, z_N) = NOT\; OR(y_1 \wedge z_1, \ldots, y_N \wedge z_N) .\]
\begin{Theorem}
\cite{ks,razborov}
\label{thm:red}
Any deterministic protocol (or nondeterministic protocol or bounded-error probabilistic protocol) for 
computing $DISJ$ requires communicating $\Omega(N)$ bits, even
if it is promised that $y_1, \ldots, y_N$ and $z_1, \ldots, z_N$ are such that there is
at most one $i:y_i=z_i=1$.
\end{Theorem}

We have $NE^d(x_1, \ldots, x_N)=NOT\; OR(x_1, \ldots, x_N)$ for inputs
$(x_1, \ldots, x_N)$ with at most one $i:x_i=1$. Hence, Theorem \ref{thm:red}
implies the second part of Theorem \ref{thm:rdw}.
\qed

\section{Conclusion and open problems}

We have shown that, for the iterated 3-bit non-equality function, $Q_E(NE^m)=O(D(NE^m)^{0.8675...})$.
This is the first example of a gap between $Q_E(f)$ and $D(f)$ that is more than a factor of 2.
We think that there are more exact quantum algorithms that are waiting to be discovered.

Some possible directions for future work are:
\begin{enumerate}
\item
Can we improve on $Q_E(NE^d)=O(2.593...^d)$, either by using our methods or in some other way?
\item
If $Q_E(NE^d)$ is asymptotically larger than $Q_2(NE^d)=\Theta(2.121...^d)$, can we prove that? 
There are cases in which we can show lower bounds on $Q_E(f)$ that are asymptotically larger than
$Q_2(f)$ \cite{Beals} but the typical lower bound methods are based on polynomial degree $deg(f)$
and, thus, are unlikely to apply to $NE^d$ for which $deg(NE^d)=2^d$ is smaller than both $Q_E(NE^d)$ and
$Q_2(NE^d)$. 
\item
Our definition of $p$-computing captures the properties that an algorithm for a
function $f$ should satisfy so that we can substitute it instead of a query to $x_i$
into Algorithm 1 for $NE(x_1, \ldots, x_N)$. Are there other definitions of 
computability that correspond to the possibility of substituting the algorithm 
instead of a query into some algorithm? 
\item
How big can the gap between $Q_E(f)$ and $D(f)$ be (for total functions $f$)? 
Currently, the best lower bound is $Q_E(f)=\Omega(D(f)^{1/3})$ by Midrij\=anis
\cite{Midrijanis}.
\item
There are other examples of functions $f^d(x_1, \ldots, x_{n^d})$ 
with $deg(f)=O(D(f)^c)$, $c<1$,
obtained by iterating different basis functions $f(x_1, \ldots, x_n)$ \cite{K,A04}.
Can we show that $Q_E(f^d)=O(D(f^d)^c)$, $c<1$ for those functions as well?
\item
Computer experiments by Montanaro et al. \cite{M+} 
show that exact quantum algorithms can be quite common:
$Q_E(f)<D(f)$ for many functions $f$ on a small number of variables. 
In particular, they indicate that $Q_E(f)<D(f)$ for many symmetric Boolean functions.
For some of those functions, exact quantum algorithms were developed by 
Ambainis et al. \cite{AI}
but, in other cases, we still do not have an explicit exact algorithm. 
\item
More generally, can we develop a general framework that
will be able to produce exact algorithms for a variety of functions $f(x_1, \ldots, x_N)$?
\item
What is $Q_E(f)$ for a random $N$-variable Boolean function $f(x_1, \ldots, x_N)$?
We know that $Q_E(f)\leq N$ (trivially) and $Q_E(f)\geq Q_2(f)=\frac{N}{2}+o(N)$ \cite{Dam,AW}. 
Is one of these two bounds optimal? 

A similar question was recently resolved for $Q_2(f)$,
by showing that $Q_2(f)=\frac{N}{2}+o(N)$ for a random $f$, with a high probability \cite{AW}.
\end{enumerate}

{\bf Acknowledgments.} 
I thank Kaspars Balodis, Aleksandrs Belovs, Ashley Montanaro, Juris Smotrovs, Ronald de Wolf, Abuzer Yakaryilmaz and the referees 
of both conference and journal versions of this paper for useful comments and Ronald 
de Wolf for pointing out Theorem \ref{thm:rdw} and its proof.

\end{document}